\documentclass[letter]{aa} 

\usepackage{graphicx}
\usepackage{placeins}
\usepackage{txfonts}
\usepackage[colorlinks=true,linkcolor=bluea,allcolors=blue]{hyperref}

\begin{document}

   \title{Absence of nuclear polycyclic aromatic hydrocarbon emission from a compact starburst: The case of the type-2 quasar Mrk\,477}


   \author{C. Ramos Almeida\inst{1,2},
          D. Esparza-Arredondo\inst{1,2},
          O. Gonz\'alez-Mart\'in\inst{3}, 
I. Garc\'ia-Bernete\inst{4},
M. Pereira-Santaella\inst{5,6},
A. Alonso-Herrero\inst{7},
J. A. Acosta-Pulido\inst{1,2},
P. S. Bessiere\inst{1,2},
N. A. Levenson\inst{8},
C. N. Tadhunter\inst{9},
D. Rigopoulou\inst{4},
M. Mart\' inez-Paredes\inst{10},
S. Cazzoli\inst{11},
\and 
B. Garc\' ia-Lorenzo\inst{1,2}
          }

   \institute{Instituto de Astrof\' isica de Canarias, Calle V\' ia L\'actea, s/n, E-38205, La Laguna, Tenerife, Spain\\
              \email{cra@iac.es}
              \and Departamento de Astrof\' isica, Universidad de La Laguna, E-38206, La Laguna, Tenerife, Spain
          \and Instituto de Radioastronom\' ia and Astrof\' isica (IRyA-UNAM), 3-72 (Xangari), 8701, Morelia, Mexico
\and Department of Physics, University of Oxford, Oxford OX1 3RH, UK
\and Centro de Astrobiolog\' ia (CSIC-INTA), Ctra. de Ajalvir, Km 4, 28850, Torrej\'on de Ardoz, Madrid, Spain
\and Observatorio Astron\'omico Nacional (OAN-IGN)-Observatorio de
Madrid, Alfonso XII, 3, 28014 Madrid, Spain
\and Centro de Astrobiolog\' ia (CAB), CSIC-INTA, Camino Bajo del Castillo s/n, E-28692, Villanueva de la Cañada, Madrid, Spain
\and Space Telescope Science Institute, Baltimore, MD 21218, USA 
\and Department of Physics \& Astronomy, University of Sheffield, S3 7RH Sheffield, UK
\and Korea Astronomy and Space Science Institute 776, Daedeokdae-ro, Yuseong-gu, Daejeon, Republic of Korea (34055)
\and Instituto de Astrof\' isica de Andaluc\' ia – CSIC, Glorieta de la Astronomía s/n, 18008 Granada, Spain
}

%

   \date{Received November 8, 2022; accepted December 1, 2022}

 
  \abstract
{Mrk\,477 is the closest type-2 quasar, at a distance of 163 Mpc. This makes it an ideal laboratory for studying the interplay between nuclear activity and star formation with a great level of detail and signal-to-noise. In this Letter we present new mid-infrared (mid-IR) imaging and spectroscopic data with an angular resolution of 0.4\arcsec~($\sim$300 pc) obtained with the Gran Telescopio Canarias instrument CanariCam. The N-band (8-13 $\mu$m) spectrum of the central $\sim$400 pc of the galaxy reveals [S IV]$\lambda$10.51 $\mu$m emission, but no 8.6 or 11.3 $\mu$m polycyclic aromatic hydrocarbon (PAH) features, which are commonly used as tracers of recent star formation. This is in stark contrast with the presence of a nuclear starburst of $\sim$300 pc in size, an age of 6 Myr, and a mass of 1.1$\times$10$^{\rm 8}$ M$_{\rm \sun}$, as constrained from ultraviolet \textit{Hubble} Space Telescope observations. Considering this, we argue that even the more resilient, neutral molecules that mainly produce the 11.3 $\mu$m PAH band are most likely being destroyed in the vicinity of the active nucleus despite the relatively large X-ray column density, log N$_{\rm H}$=23.5 cm$^{-2}$, and modest X-ray luminosity, 1.5$\times$10$^{43}$ erg~s$^{\rm -1}$. This highlights the importance of being cautious when using PAH features as star formation tracers in the central region of galaxies to evaluate the impact of feedback from active galactic nuclei.

}

   
\titlerunning{Absence of nuclear PAH emission from a compact starburst}
\authorrunning{C. Ramos Almeida et al.}

   \maketitle
%

\section{Introduction}
\label{intro}

Type-2 quasars (QSO2s) are optically selected active galactic nuclei (AGN) with L$_{\rm [OIII]}>$10$^{8.5}L_{\sun}\sim$M$_B<$-23 that show permitted emission lines with a full width at half maximum (FWHM) of $<$2000 km s$^{-1}$ \citep{2008AJ....136.2373R}. They might constitute a key evolutionary phase when the AGN is clearing up gas and dust to eventually shine as an unobscured quasar \citep{1988ApJ...325...74S,2009ApJ...696..891H}. Mrk\,477 (SDSS J144038.1+533016, I Zw 92) is the closest QSO2 in the Quasar Feedback (\href{http://research.iac.es/galeria/cra/qsofeed/}{QSOFEED}) sample \citep{2022A&A...658A.155R}, having a redshift of z=0.037 and a luminosity distance D$_L$=163 Mpc (scale of 735 pc/\arcsec). It is indeed the nearest obscured quasar \citep{2015MNRAS.454..439V}, and it shows broad lines in the polarized flux spectrum \citep{1992ApJ...397..452T}. Its [O III] luminosity, 10$^{\rm 8.94}$ L$_{\sun}$, corresponds to a bolometric luminosity of 1.3$\times$10$^{\rm 46}$ erg~s$^{-1}$ using a correction factor of 3500 \citep{2004ApJ...613..109H}. A more conservative estimate of L$_{\rm bol}$=1.8$\times$10$^{\rm 45}$ erg~s$^{-1}$ was derived by \citet{2021MNRAS.500.1491T} using the extinction-corrected [O III] luminosity and the correction factor of 454 from \citet{2009A&A...504...73L}. The black hole mass and Eddington ratio reported by \citet{2018ApJ...859..116K} for this QSO2 are log M$_{\rm BH}$=7.16$\pm$0.82 M$_{\sun}$ and L$_{\rm bol}$/L$_{\rm Edd}$=0.41$\pm$0.83. The QSO2 is radio-quiet, but it has a 1.4 GHz luminosity of 1.9$\times$10$^{\rm 23}$ W~Hz$^{-1}$, and 8.4 GHz Very Large Array (VLA) data at 0.26\arcsec~resolution show a triple radio source of $\sim$1.3\arcsec~(1 kpc) in size and a position angle (PA) of $\sim$30--35\degr. This orientation is the same as that of the [O III] emission observed with the \textit{Hubble} Space Telescope \citep[HST;][]{1997ApJ...482..114H,2018ApJ...856..102F}, which has a total extent of $\sim$5\arcsec~(3.7 kpc). 
Data from the Space Telescope Imaging Spectrograph (STIS) revealed a compact ionized outflow (r=0.54 kpc), 
with a maximum velocity of -500 km~s$^{-1}$ \citep{2018ApJ...856..102F}.

  \begin{figure*}
   \centering
   {\par\includegraphics[width=0.802\columnwidth]{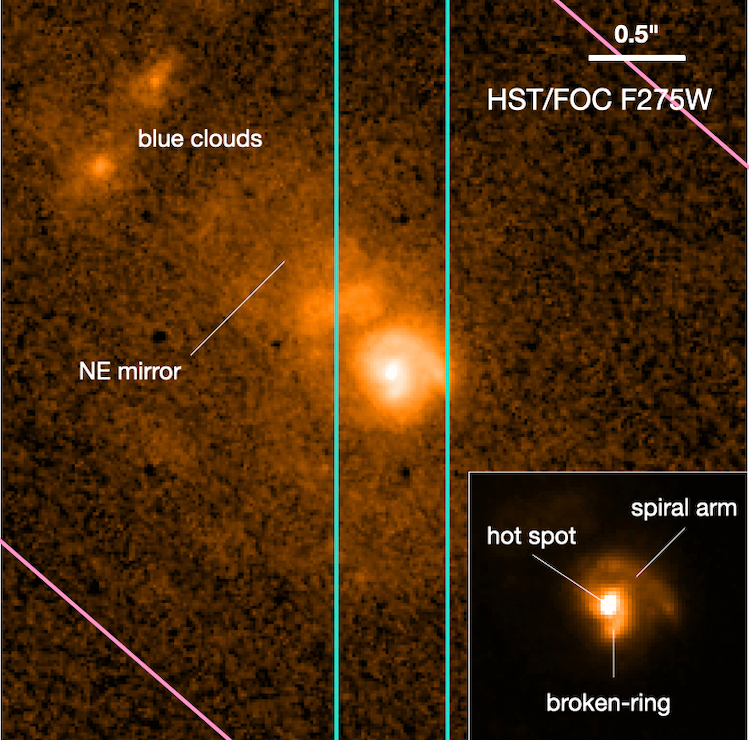}
   \includegraphics[width=0.87\columnwidth]{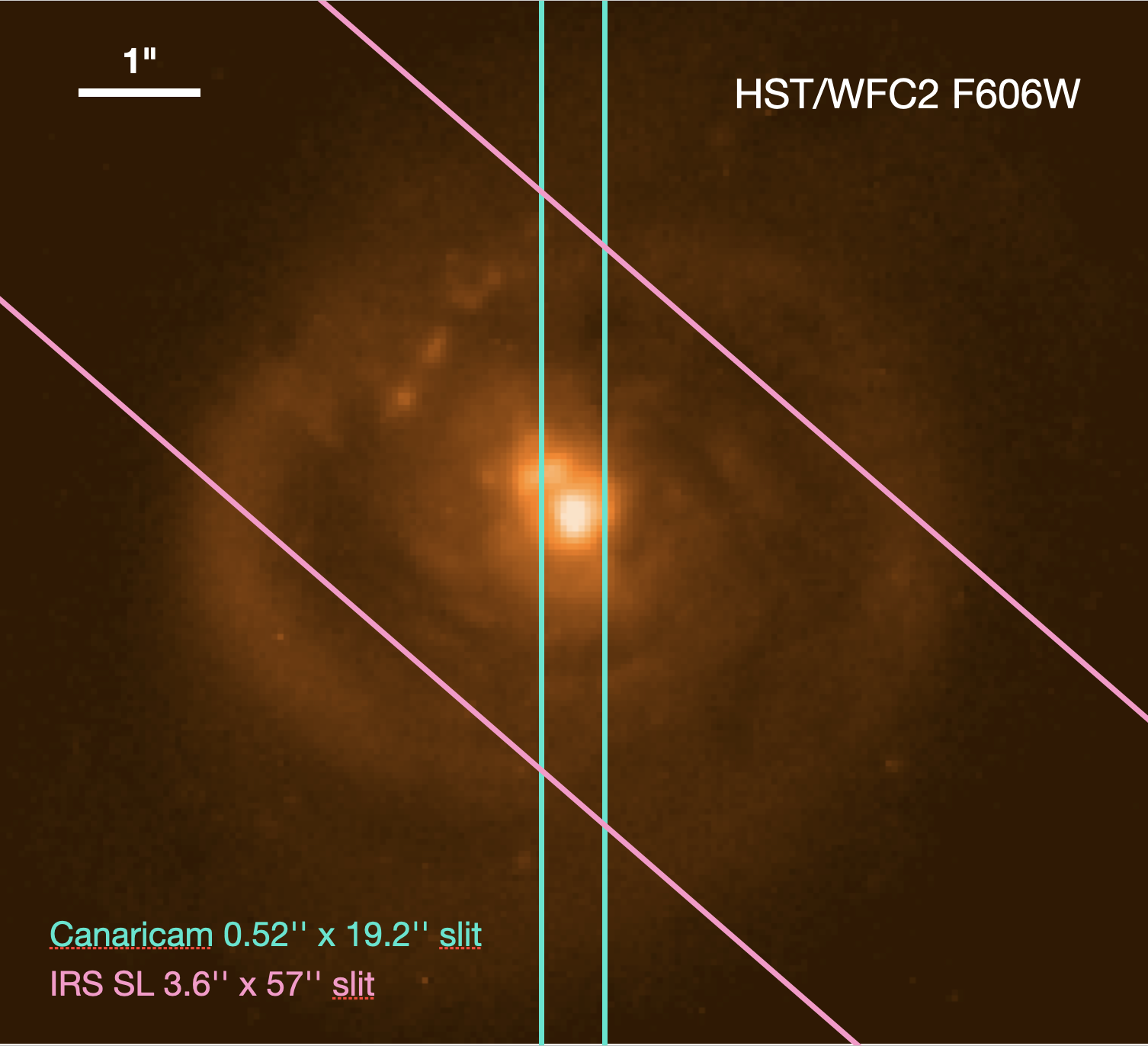}\par}
   \caption{UV and optical HST images of Mrk\,477. The Faint Object Camera (FOC) F275W image (ID: 6702; PI: Kay) of the central 3.5\arcsec$\times$3.5\arcsec~of the galaxy is shown in the left panel, which includes an inset of the central 1\arcsec$\times$1\arcsec~region (735$\times$735 pc$^2$). The Wide Field and Planetary Camera 2 (WFPC2) F606W image (ID: 8597; PI: Regan) of the central 9.5\arcsec$\times$8.6\arcsec~is shown in the right panel. The CanariCam (0.52\arcsec$\times$19.2\arcsec, PA=0\degr) and IRS SL slits (3.6\arcsec$\times$57\arcsec, PA=49\degr) are shown as solid cyan and pink lines. North is up and east to the left.}
              \label{fig1}%
    \end{figure*}

The host galaxy is a spiral of stellar mass log M$_*$=10.7$\pm$0.1 M$_{\sun}$, calculated from the 2MASS Extended Source Catalog K-band magnitude, as described in \citet{2022A&A...658A.155R}, with a major axis PA$\sim$112\degr~and inclination of 22\degr~\citep{2018ApJ...856..102F}. It is clearly interacting with a companion galaxy $\sim$36 kpc to the northeast \citep{2015MNRAS.454..439V}. 
On smaller scales, ultraviolet (UV) and optical images obtained with the HST, shown in Fig.~\ref{fig1}, reveal a complex nuclear region, of $\sim$0.4\arcsec~(300 pc) in diameter, first reported by \citet{1997ApJ...482..114H}. The inset in the left panel of Fig.~\ref{fig1} shows its three main components: the hot spot, the broken-ring, and the small spiral arm. At $\sim$0.6\arcsec~(440 pc) from the hot spot in the northeast direction, there is an off-nuclear scattering region named the ``NE mirror'' by \citet{2002ApJ...567..790K}. These authors analyzed UV imaging polarimetry data and reported that the hot spot radiation is slightly polarized, but not dominated by scattered light, and that there has to be an additional continuum source outside the broad line region to explain the observations. \citet{1997ApJ...482..114H} used an HST far-UV spectrum of the central 1.74\arcsec$\times$1.74\arcsec~region to study the underlying stellar populations of this QSO2. This region includes the hot spot, the broken-ring, the small spiral arm, and the NE mirror (see Fig. \ref{fig1}). They reported the detection of Si III$\lambda$1417 and C III$\lambda\lambda$1426,1428 photospheric absorption lines, which are strong in late O and early B supergiants. The UV spectrum, in combination with an HST UV continuum image and ground-based optical and near-infrared (near-IR) spectra, revealed the presence of a compact starburst, with an age of $\sim$6 Myr, a duration of 1 Myr, and solar or twice-solar metallicity. The light of this starburst, for which \citet{1997ApJ...482..114H} calculated a bolometric luminosity of 3.2$\times$10$^{\rm 10}$ L$_{\sun}$ and a mass of 1.1$\times$10$^8$ M$_{\rm \sun}$, 
dominates the observed continuum from the UV to the near-IR. This is most likely the additional continuum source proposed by \citet{2002ApJ...567..790K} to explain the imaging polarimetry observations. Using HST UV images of the central region in two filters (F275W and F342W), \citet{2002ApJ...567..790K} reported bluer colors for the small spiral arm than for the rest of features shown in the inset of Fig.~\ref{fig1}, and claimed that this might be the location of the starburst.  

The presence of this compact starburst was the main motivation for observing Mrk\,477 with the 10.4 Gran Telescopio Canarias (GTC) instrument CanariCam \citep{2003SPIE.4841..913T,2005RMxAC..24....7P}. This mid-IR camera and spectrograph provides diffraction-limited data, permitting us to obtain spectra of the inner $\sim$400 pc of the QSO2. The UV radiation from the young O and B stars in this starburst should excite polycyclic aromatic hydrocarbon (PAH) features in the mid-IR, which CanariCam can study with unprecedented spatial resolution in a QSO2. These features are particularly bright in regions illuminated by UV-bright,    early-type stars (e.g., H II regions and reflection nebulae; \citealt{2004ApJ...613..986P}), and they are commonly used as tracers of star formation \citep{1998ApJ...498..579G,1999AJ....118.2625R}. Evidence for star formation based on the detection of PAH features has been reported for AGN of different luminosities, from kiloparsec scales \citep{2008AJ....136.1607Z,2016MNRAS.455.4191Z,2009ApJ...705...14D,2012ApJ...746..168D} to as close as tens of parsecs from the AGN \citep{2013MNRAS.429.2634S,2014ApJ...780...86E,2018ApJ...859..124E,2019ApJ...871..190M}. However, it has also been demonstrated that AGN radiation and/or shocks suppress the short-wavelength features (6.2, 7.7, and 8.6 $\mu$m) by modifying the structure of the aromatic molecules or destroying the smallest grains \citep{2007ApJ...656..770S,2010ApJ...724..140D}. On the other hand, the more resilient, larger, and neutral molecules that produce the 11.3 and 17 $\mu$m features are enhanced in AGN relative to the short-wavelength PAHs \citep{2022A&A...666L...5G,2022MNRAS.509.4256G}, and they can still be used to determine  star formation rates (SFRs) in AGN with bolometric luminosities $\la$10$^{\rm 46}$ erg~s$^{-1}$ \citep{2022ApJ...925..218X}. Investigating the suitability of PAHs as star formation tracers in the innermost regions of AGN (i.e., the central kiloparsec) is important because that is where we can evaluate the impact of AGN feedback on star formation, given the short timescales of nuclear activity \citep{2014ApJ...782....9H}. Other star formation tracers, such as H$\alpha$, [Ne II], the 24 $\mu$m flux, and even the far-IR luminosities, contain some degree of AGN contamination. The PAH features are considered to be ``cleaner'' probes of recent star formation, although it has been argued that they could also be excited by AGN radiation \citep{2017MNRAS.470.3071J}.

Here we assume a cosmology with H$_0$=70 km~s$^{-1}$ Mpc$^{-1}$, $\Omega_m$=0.3, and $\Omega_{\Lambda}$=0.7. The measurements from other works discussed here have been converted to this cosmology.

\section{Observations and data reduction}
\label{observations}

Mrk\,477 was observed on March 5, 2020, under photometric conditions, with CanariCam. This mid-IR instrument, relocated in 2019 to the folded Cassegrain E focus of the 10.4 GTC, uses a Raytheon 320$\times$240 Si:As detector, which covers a field of view of 26\arcsec~$\times$19\arcsec~and has a pixel size of 0.0798\arcsec. A chopping–nodding technique was used to remove the time-variable sky background, the thermal emission from the telescope, and the detector noise. Chopping and nodding throws were 15\arcsec, with chop and nod PAs of 90 and 270$\degr$, respectively.
The observations were done in queue mode, as part of program GTC83-19B (PI: Ramos Almeida). The image observations were taken in the narrow Si-2 filter ($\lambda_c$=8.7 $\mu$m, $\Delta\lambda$=1.1 $\mu$m) and consisted of two exposures of 417 s each, which we combined after they were reduced to produce a single image of 834 s on-source. The airmass during the imaging observations was 1.1. The point spread function (PSF) standard star HD 128000 was observed after the science target in the same filter to accurately sample the image quality and allow flux calibration. We reduced the data using the {REDCAN} pipeline \citep{2013A&A...553A..35G}, which, in the case of the imaging, performs sky subtraction, stacking of the individual images, rejection of bad frames, and flux calibration.

For the spectroscopic observations, we used the low spectral resolution N-band grating, which has a nominal resolution R = $\lambda$/$\Delta\lambda$=175 and covers the spectral range 7.5–13 $\mu$m. We used a slit width of 0.52\arcsec, oriented with a PA=0$\degr$ (see Fig.~\ref{fig1}). We 
integrated for 1296 s of total on-source time, split into two exposures of 648 s each, at an airmass of 1.1. Immediately after, we obtained a spectrum of the star HD 128000 to provide flux calibration and telluric and slit-loss corrections. The data reduction with {REDCAN} consists of sky subtraction, stacking of individual observations, rejection of bad frames, wavelength calibration, trace determination, spectral extraction, flux calibration, and combination into a single spectrum. The spectral extraction can be done either as a point source (i.e., using an aperture that increases with wavelength to account for the decreasing angular resolution and performing a slit-loss correction) or as an extended source, in which case a fixed aperture is used and no slit-loss corrections are applied. Here we focus on the nuclear spectrum, extracted as a point source (average aperture size of 0.5\arcsec$\sim$400 pc), but we also consider two additional spectra extracted as extended sources, in apertures of 1.28\arcsec~and 1.60\arcsec~(940 and 1175 pc). The errors were estimated as the sum of the statistical error (i.e., squared root of the number of counts) and 15\% of the flux calibration uncertainty \citep{2016MNRAS.455..563A}. 




We also retrieved a mid-IR spectrum of Mrk\,477 from the Combined Atlas of Sources with \textit{Spitzer} IRS Spectra (\href{https://cassis.sirtf.com/atlas/welcome.shtml}{CASSIS}; AORkey: 17643008), obtained with the Infrared Spectrograph (IRS) of the \textit{Spitzer} Space Telescope. The observations were done on June 30, 2006 (PI: G. Rieke, program ID: 30443), using the Short-Low (SL) and Long-Low (LL) spectrograph modules, which cover the wavelength range 5.3--38 $\mu$m with a spectral resolution of R = $\lambda$/$\Delta\lambda\sim$60--130. The SL1 and SL2 slits have widths of 3.6 and 3.7\arcsec~(2.7 kpc), and they were oriented with a PA=49\degr~(see Fig.~\ref{fig1}). The LL slits have widths of 10.5 and 10.7\arcsec~(7.8 kpc) and PA=-35\degr. CASSIS identifies Mrk\,477 as point-like, and therefore, optimal extraction produces the best flux-calibrated spectrum.
Finally, we retrieved \textit{Spitzer} Infrared Array Camera (IRAC) images in the 5.8 and 8 $\mu$m arrays (IRAC3 and IRAC4) from the same program using the \textit{Spitzer} Heritage Archive. The data were taken on July 6, 2006, and the nominal PSF values in IRAC3 and IRAC4 are 1.88\arcsec~and 1.98\arcsec. 

\section{Results}
\label{results}

\subsection{Mid-IR continuum}
\label{continuum}

Using the CanariCam Si-2 images, we measured an angular resolution of 0.39\arcsec~(290 pc) from the FWHM of the PSF star. Mrk\,477 is unresolved in the mid-IR, since we measure FWHMs of 0.38 and 0.41\arcsec~in the individual images and 0.40\arcsec~in the combined image (295 pc). 

\begin{figure*}
   \centering
  \includegraphics[width=1.4\columnwidth]{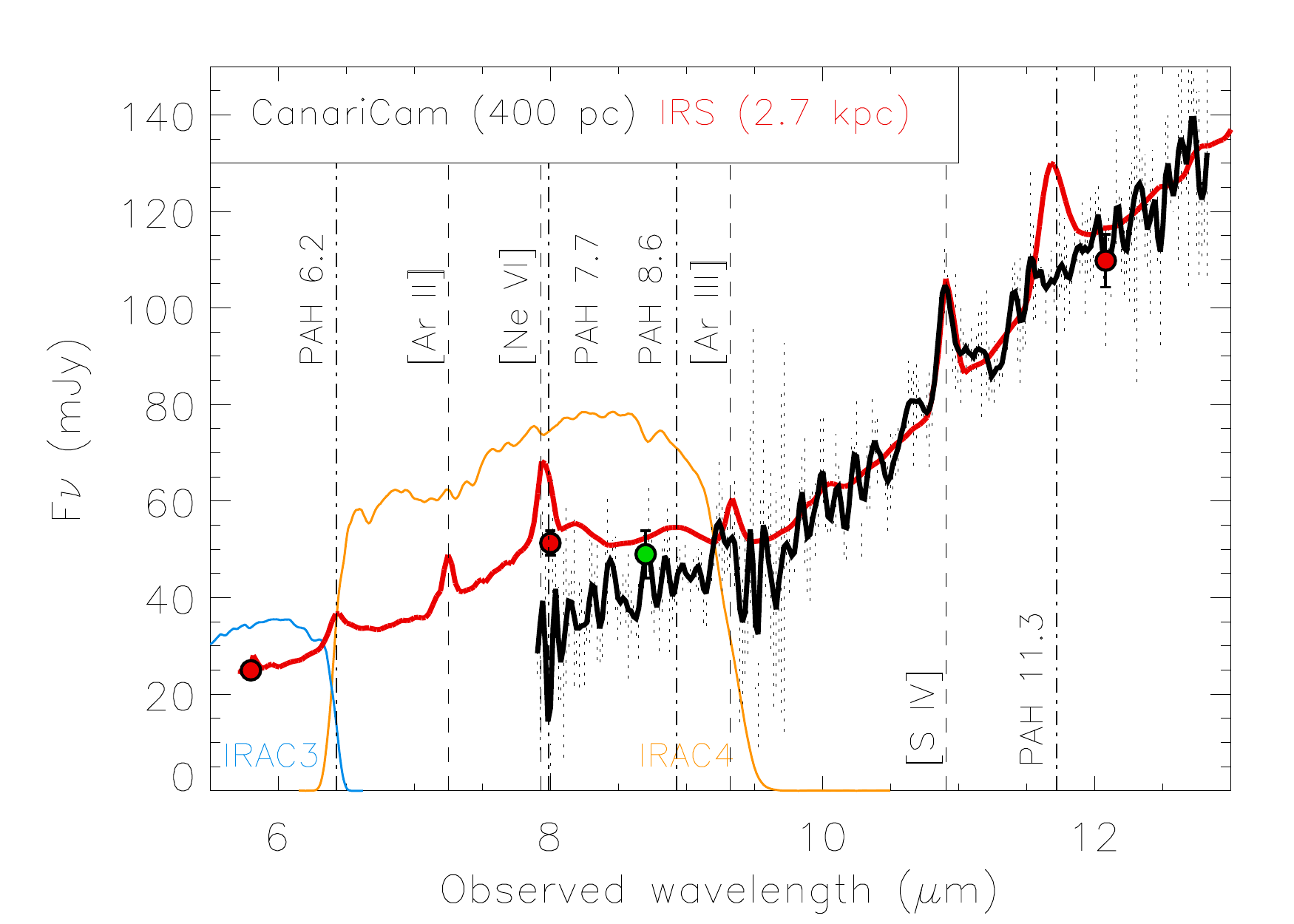}
   \caption{GTC/CanariCam nuclear spectrum of Mrk\,477 (0.5\arcsec$\sim$400 pc; solid black line, with errors indicated as dotted lines). Weak silicate absorption and [S IV]$\lambda$10.51 $\mu$m emission are the only spectral features detected. The solid red line corresponds to the IRS spectrum (3.6\arcsec$\sim$2.7 kpc), which,  in addition to the ionized and warm molecular hydrogen emission lines, shows clear PAH emission bands at 6.2, 7.7, 8.6, and 11.3 $\mu$m. Filled dots are the fluxes obtained from images in the CanariCam Si-2 filter (in green), IRAC 5.8 and 8 $\mu$m, and WISE 12 $\mu$m bands (in red). The transmission curves of the IRAC3 and IRAC4 filters are shown as solid blue and orange lines.}
              \label{fig2}%
    \end{figure*}

In Fig.~\ref{fig2} we show the CanariCam nuclear spectrum of Mrk\,477. 
There is good agreement between the flux calibration of the nuclear spectrum and the 8.7 $\mu$m flux obtained from the image in the Si-2 filter (green dot in Fig.~\ref{fig2}), which is 49$\pm$5 mJy. This is expected considering the excellent conditions during the observations (seeing size and stability and photometric night). 
The continuum shows a positive slope, going from $\sim$35 mJy at 8 $\mu$m to $\sim$130 mJy at 12.5 $\mu$m. This continuum shape resembles those of Seyfert 2 galaxies with shallow silicate absorption features \citep{2014MNRAS.443.2766A,2014MNRAS.439.3847R}. Shallow silicate absorption is generally observed in IRS spectra of QSO2s, with S$_{\rm 9.7}$ values ranging from 0 to -0.5 \citep{2008AJ....136.1607Z}. For Mrk\,477, \citet{2016MNRAS.455.4191Z} reported S$_{\rm 9.7}$=0.31, as measured from the IRS spectrum.

From Fig.~\ref{fig2}, the excellent match between the continua of the IRS and CanariCam spectra from $\sim$10 $\mu$m redward is evident, and the slightly different spectral shape at $\lambda$<10 $\mu$m is due to the presence of broad 7.7 and 8.6 $\mu$m PAH features in the IRS spectrum. This similarity in continuum shape is in principle unexpected considering the different spatial scales probed by the two spectra (inner 400 pc versus 2.7 kpc), but Mrk\,477 is compact in the mid-IR, as measured from the CanariCam and IRAC images. The FWHMs that we measure in the IRAC 5.8 and 8 $\mu$m images, 1.85\arcsec~and 2.02\arcsec, are much smaller than the slit widths of the SL IRS module and the IRS PSF in the blue filter ($\sim$15 $\mu$m), which is 3.8\arcsec. The compactness of the mid-IR emission explains the excellent agreement between the IRAC and Wide-field Infrared Survey Explorer (WISE) photometry and the IRS spectrum, $\la$5\% (see Fig.~\ref{fig2}), even when large apertures are considered (up to 5.8\arcsec~in the case of IRAC 5.8 and 8 $\mu$m, and 22\arcsec~for WISE 12 $\mu$m). The IRAC and WISE fluxes were retrieved from the NASA/IPAC Infrared Science Archive (Caltech/JPL), and the associated uncertainties are smaller than the 5\% error bars plotted in Fig.~\ref{fig2}. Considering this, and the similarity between the CanariCam spectrum and those of Seyfert 2 galaxies with shallow absorption (e.g., NGC\,3081; \citealt{2014MNRAS.439.3847R}), we conclude that the bulk of the mid-IR emission in Mrk\,477 must be dominated by the AGN continuum. This was already noted by \citet{1997ApJ...482..114H}, based on the shape of the IR spectral energy distribution. Dust heated by star formation is usually at lower temperatures and produces steeper slopes in the mid-IR \citep{2011MNRAS.414..500H}. Indeed, if we decompose the IRS spectrum using a linear combination of empirical templates, we find the AGN contribution to the 5--22 $\mu$m continuum to be 93\% (see Appendix \ref{appendix}).



\subsection{Mid-IR emission lines}
\label{emission}

In the spectral range shown in Fig.~\ref{fig2}, the IRS spectrum shows [S IV]$\lambda$10.51 $\mu$m, [Ar III]$\lambda$8.99 $\mu$m, [Ne VI]$\lambda$7.65 $\mu$m, and [Ar II]$\lambda$6.99 $\mu$m, 
as well as the PAH features at 6.2, 7.7, 8.6, and 11.3 $\mu$m. These PAH bands were studied in \citet{2009ApJ...705...14D}, and the authors reported a deficit of 6.2 and 7.7 $\mu$m emission in Mrk\,477 as compared to the 11.3 and 17 $\mu$m bands, once they subtracted a starburst template from the IRS spectrum. 
Shocks may be responsible for the unusual PAH band ratios seen over large scales in Mrk\,477, since they induce a rapid increase in temperature and density that affects the chemical and physical properties of both the gas and solid phases of the interstellar medium \citep{2021MNRAS.504.5287R}.

In contrast with the IRS spectrum, we do not see any evidence of PAH emission in the nuclear CanariCam spectrum. In particular, we do not detect the 11.3 $\mu$m PAH feature, which is mainly produced by larger and thus more resilient neutral molecules that can survive in the inner tens or hundreds of parsecs of nearby Seyfert galaxies and Palomar-Green (PG) quasars (\citealt{2014ApJ...780...86E,2019ApJ...871..190M}). The only spectral features detected in the nuclear CanariCam spectrum are the shallow 10 $\mu$m silicate absorption band and the [S IV]$\lambda$10.51 $\mu$m emission line (IP=47 eV). This emission line is clearly detected both in the CanariCam and IRS spectra, with practically the same luminosity and equivalent width (EW; see Table \ref{tab1}). This indicates that, whilst the continuum and the [S IV] emission are mostly nuclear ($\la$400 pc; i.e., the region that includes the hot spot, broken-ring, spiral arm, and part of the NE mirror, as shown in Fig.~\ref{fig1}), the PAH emission detected in the IRS spectrum is coming from exposed photodissociation regions (PDRs), reflection nebulae, and/or H II regions on larger scales. We do not detect PAH emission in the CanariCam spectra extracted in apertures of 1.28\arcsec~and 1.60\arcsec~either (see Table \ref{tab1}). Integrating beyond these regions only introduces noise, since we do not detect extended emission in either the CanariCam images or spectrum. 
The IRS SL slit, much wider than CanariCam's and with a different orientation, also includes the NE mirror, the ``blue clouds'' in the left panel of Fig.~\ref{fig1}, which have even bluer UV colors than the small spiral arm \citep{2002ApJ...567..790K}, and part of the large-scale galaxy spiral arms shown in the right panel of Fig.~\ref{fig1}.


\begin{table*}
\centering
\begin{tabular}{lccccccc}
\hline
\hline
Instrument & \multicolumn{2}{c}{Aperture size}   &  \multicolumn{2}{c}{[S IV]$\lambda$10.51 $\mu$m}   & \multicolumn{3}{c}{PAH 11.3 $\mu$m}   \\
&     & & L$_{\rm 10.51~\mu m}$ & EW                     &  L$_{\rm 11.3~\mu m}$ & EW&  SFR                    \\
&(\arcsec) & (pc) &  (10$^{\rm 41}$ erg~s$^{\rm -1}$) & ($\mu$m)  &   (10$^{\rm 41}$ erg~s$^{\rm -1}$) & ($\mu$m) & (M$_{\rm \sun}$~yr$^{\rm -1}$) \\
\hline
CanariCam &  0.50 &  400 & 3.10$\pm$0.04 & 0.033$\pm$0.001 & <2.05 & <0.017$\pm$0.001 & <0.45 \\
CanariCam & 1.28 &  940 & 3.01$\pm$0.04 & 0.031$\pm$0.001 & <2.37 & <0.016$\pm$0.001 & <0.52 \\
CanariCam & 1.60 & 1175 & 2.88$\pm$0.04 & 0.029$\pm$0.001 & <2.36 & <0.020$\pm$0.001 & <0.52 \\
IRS       & 3.60 & 2650 & 2.70$\pm$0.04 & 0.029$\pm$0.001 & 10.1$\pm$0.1$^*$ & 0.049$\pm$0.001 & 2.41$\pm$1.01 \\
\hline
\end{tabular}
\caption{[S IV] and 11.3 $\mu$m PAH measurements from the CanariCam nuclear and extended source spectra, and the IRS spectrum. Line fluxes were integrated over a local continuum in the rest-frame ranges [10.3,10.7] and [11.1,11.6] $\mu$m. This continuum was determined via linear interpolation of the average flux in two narrow bands on both sides of each feature ([10.0,10.3] and [10.7,11.0] $\mu$m for [S IV], and [10.7,11.1] and [11.6,12.0] $\mu$m for the PAH). *The PAH luminosity includes a multiplicative factor of two to make it comparable with PAHFIT-derived measurements \citep{2007ApJ...656..770S}. Upper limits at a 2$\sigma$ significance
are indicated with the < symbol for non-detections, where $\sigma$ is the average of the continuum noise calculated in the two bands adjacent to the PAH. SFRs were calculated from L$_{\rm 11.3~\mu m}$ following Eq.~12 in \citet{2016ApJ...818...60S} and converted from the Kroupa to Salpeter initial mass function by dividing them by 0.67. Errors were measured using 150 Monte Carlo simulations.}
\label{tab1} 
\end{table*}

Further evidence for this scenario comes from the sizes measured from the IRAC3 and IRAC4 images. 
By fitting a 2D Gaussian convolved with the IRAC PSF in each filter, we derived FWHM sizes of 1.26\arcsec$\times$0.95\arcsec~at 5.8 $\mu$m and 1.30\arcsec$\times$1.15\arcsec~at 8 $\mu$m. The slightly more extended emission that we detect at 8 $\mu$m (IRAC4) might be coming from the PAH features lying within this filter, shown in Fig.~\ref{fig2}, since the 5.8 $\mu$m filter (IRAC3) does not contain any of the strong PAH features. 

  
\section{Discussion}
\label{discussion}

In another AGN, the lack of 11.3 $\mu$m PAH emission in the nuclear spectrum could be interpreted as evidence for AGN feedback quenching star formation, but in Mrk\,477 we know from the detailed analysis presented in \citet{1997ApJ...482..114H} that this is not the case. The O and B stars in the nuclear starburst should be producing PAH emission. On this basis, we estimate upper limits for the 11.3 $\mu$m luminosity at a 2$\sigma$ significance, as in \citet{2014ApJ...780...86E}, and from them we measure SFRs<0.5 M$_{\rm \sun}$~yr$^{\rm -1}$ (see Table \ref{tab1}). This upper limit needs to be compared with the SFR that we would expect from the starburst. To do so, we used Starburst99 (v7.0.1; \citealt{1999ApJS..123....3L}) to model an instantaneous burst of star formation of mass, age, and duration of 1.1$\times$10$^{\rm 8}$ M$_{\rm \sun}$, 6 Myr, and $\sim$1 Myr \citep{1997ApJ...482..114H}, as well as a continuous SFR of 1 M$_{\rm \sun}$~yr$^{\rm -1}$. We then integrated the UV luminosity of the two models between 912 and 3000 \AA, which is the UV radiation that excites the PAH molecules \citep{2001ApJ...554..778L}, and we find

\begin{equation}
    L_{\rm 912-3000~\AA}^{\rm burst} = 6.1\times L_{\rm 912-3000~\AA}^{\rm 1~M_{\rm \sun}~yr^{\rm -1}}.
\end{equation}

Thus, the nuclear starburst has a UV radiation equivalent to that of a continuous SFR=6.1 M$_{\rm \sun}$~yr$^{\rm -1}$. This SFR is of the same order as that derived from stellar population modeling of the optical SDSS spectrum (3\arcsec~diameter; SFR$\simeq$10 M$_{\rm \sun}$~yr$^{\rm -1}$ using the Salpeter initial mass function; {\color{blue} Bessiere et al., in prep.}). This is expected, since the model of the optical spectrum includes a young stellar population that is consistent with the results reported by \citet{1997ApJ...482..114H}. This continuous SFR of 6.1 M$_{\rm \sun}$~yr$^{\rm -1}$ corresponds to a 11.3 $\mu$m luminosity of 2.4$\times$10$^{\rm 42}$ erg~s$^{\rm -1}$, following Eq.~12 in \citet{2016ApJ...818...60S}, which is more than ten times higher than the upper limits that we measure from the CanariCam spectra (see Table \ref{tab1}). Therefore, the most likely explanation
for the lack of PAH emission is that even the larger molecules are being destroyed by the quasar radiation and/or shocks. Indeed, recent results based on observations from the JWST Medium-Resolution Spectrometer (MRS) of the Mid-IR Instrument (MIRI) of three nearby Seyfert galaxies provided evidence that nuclear activity has a significant impact on the ionisation state and size of the PAH grains on nuclear scales \citep{2022A&A...666L...5G}. 

Another possibility could be that the nuclear starburst is extremely obscured, and therefore, the PAH-emitting regions are buried inside optically thick layers of dust. We discard this scenario because the intrinsic extinction measured at 2150 \AA~is 0.9$\pm$0.3 mag \citep{1997ApJ...482..114H}, and the silicate feature detected in the CanariCam spectrum is weak. Dilution by the strong AGN continuum \citep{2014MNRAS.443.2766A} could also explain the lack of nuclear PAH emission. However, considering the expected PAH luminosity that we estimate from the starburst and the 2$\sigma$ upper limits that we measure from the adjacent mid-IR continuum, we favor PAH destruction over dilution to explain the lack of 11.3 $\mu$m emission in the nuclear spectrum. 
To further explore the potential destruction of PAH molecules in Mrk\,477, we need measurements of its X-ray luminosity and column density. This QSO2 was proposed as a Compton-thick candidate by \citet{1999ApJS..121..473B} and \citet{2007ApJ...657..167S}, but more recently, using data from the Nuclear Spectroscopic Telescope Array (\textit{NuSTAR}) and the X-ray Multi-Mirror Mission (\textit{XMM-Newton}), \citet{2018ApJ...854...49M} reported a column density of N$_{\rm H}$=(3.2$\pm$0.4)$\times$10$^{23}$ cm$^{\rm -2}$ and an intrinsic 2-10 keV luminosity of (1.54$\pm$0.05)$\times$10$^{43}$ erg~s$^{\rm -1}$. Using a bolometric correction factor of 20, this corresponds to L$_{\rm bol}$=3.1$\times$10$^{44}$ erg~s$^{\rm -1}$, which is much lower than those estimated from the [O III] luminosity (see Sect. \ref{intro}). Using a correction factor of 70, more appropriate for heavily obscured AGN \citep{2012ApJ...748..130M}, yields  L$_{\rm bol}$=1.1$\times$10$^{45}$ erg~s$^{\rm -1}$, which is in better agreement with the bolometric luminosity estimated from the extinction-corrected [O III] luminosity (1.8$\times$10$^{45}$ erg~s$^{\rm -1}$; \citealt{2021MNRAS.500.1491T}). 

At the X-ray luminosity of Mrk\,477, according to the theoretical predictions from \citet{1992MNRAS.258..841V} and \citet{1994ApJ...425L..37M}, the neutral molecules responsible for the 11.3 $\mu$m PAH feature should be shielded from AGN radiation up to distances as close as 150 pc from the AGN when N$_{\rm H}\geq$10$^{\rm 23}$ cm$^{\rm -2}$ (see Fig.~13 in \citealt{2014MNRAS.443.2766A}). However, more recent predictions by \citet{2019MNRAS.488..451M}, based on experimental mass spectrometry,  showed that, even considering X-ray optical depths typical of the dusty torus, the molecules' half-lives are not long enough to account for PAH detection in AGN.
From an observational point of view, \citet{2019ApJ...871..190M} reported the detection of 11.3 $\mu$m features in ground-based N-band spectra of 5 out of 13 PG quasars at z<0.1 with X-ray luminosities ranging from 2$\times$10$^{43}$ to 5$\times$10$^{44}$ erg~s$^{\rm -1}$. The spatial scales probed by their nuclear spectra are the inner 700 pc or less, showing that in some cases it is possible to shield the neutral molecules at these high AGN luminosities. 
Nevertheless, having a high column density does not necessarily imply that 11.3 $\mu$m emission is detected in the nuclear region, as shown by \citet{2020A&A...639A..43A} using mid-IR and submillimeter observations of nearby AGN.
New models and targeted high angular resolution spectroscopic observations are required for a better understanding of the diversity of PAH line ratios observed in AGN of different luminosities and obscuration levels. 

Mrk\,477 is a perfect laboratory for studying the reliability of PAHs as star formation tracers as a function of distance from the active nucleus. 
It is a luminous AGN with a high hydrogen column density, co-existing with a nuclear starburst of $\sim$300 pc in size, an age of 6 Myr, and a mass of 1.1$\times$10$^{\rm 8}$ M$_{\rm \sun}$. This starburst dominates the continuum emission from the UV to the near-IR, but does not show detectable PAH emission in the mid-IR nuclear spectrum ($\sim$400 pc). This implies that the nuclear molecules responsible for the 11.3 $\mu$m emission are most likely being destroyed by AGN radiation and/or shocks. On the other hand, the PAH features that we detect in the IRS spectrum must come from extended regions beyond the central 400 pc. This study, which has been possible thanks to the excellent observing conditions and good angular resolution of the mid-IR GTC/CanariCam data, can be improved, in terms of sensitivity and wavelength coverage, with the integral field mode capabilities of the JWST.


\begin{acknowledgements}
Based on observations made with the GTC, installed at the Spanish Observatorio del Roque de los Muchachos of the IAC, on the island of La Palma. CASSIS is a product of the IRS instrument team, supported by NASA and JPL, by the ``Programme National de Physique Stellaire'' (PNPS) of CNRS/INSU co-funded by CEA and CNES and through the ``Programme National Physique et Chimie du Milieu Interstellaire'' (PCMI) of CNRS/INSU with INC/INP co-funded by CEA and CNES.
CRA acknowledges the hospitality of the Kavli Institute for Cosmology of the University of Cambridge, where this manuscript was written, in August 2022. This stay was funded by the Spanish MICINN through the Spanish State Research Agency, under Severo Ochoa Centres of Excellence Programme 2020-2023 (CEX2019-000920-S).  
CRA and PSB acknowledge support from the projects ``Quantifying the impact of quasar feedback on galaxy evolution'', with reference EUR2020-112266, funded by MICINN-AEI/10.13039/501100011033 and the European Union NextGenerationEU/PRTR, and from the Consejer\' ia de Econom\' ia, Conocimiento y Empleo del Gobierno de 
Canarias and the European Regional Development Fund (ERDF) under grant ``Quasar feedback and molecular gas reservoirs'', with reference ProID2020010105, ACCISI/FEDER, UE. AAH acknowledges financial support from grant PID2021-124665NB-I00 funded by the Spanish Ministry of Science and Innovation. SC acknowledges financial support from the State Agency for Research of
the Spanish MCIU through the ``Center of Excellence Severo Ochoa'' award to the IAA (SEV-2017-0709). We finally thank the referee for his constructive report, and the GTC staff for their constant support. 
\end{acknowledgements}

\bibliographystyle{aa} 
\bibliography{aanda} 

\begin{appendix} 
\section{Spectral decomposition of the IRS spectrum}\label{appendix}

\begin{figure}[!h]
  \centering
 \includegraphics[width=1.0\columnwidth]{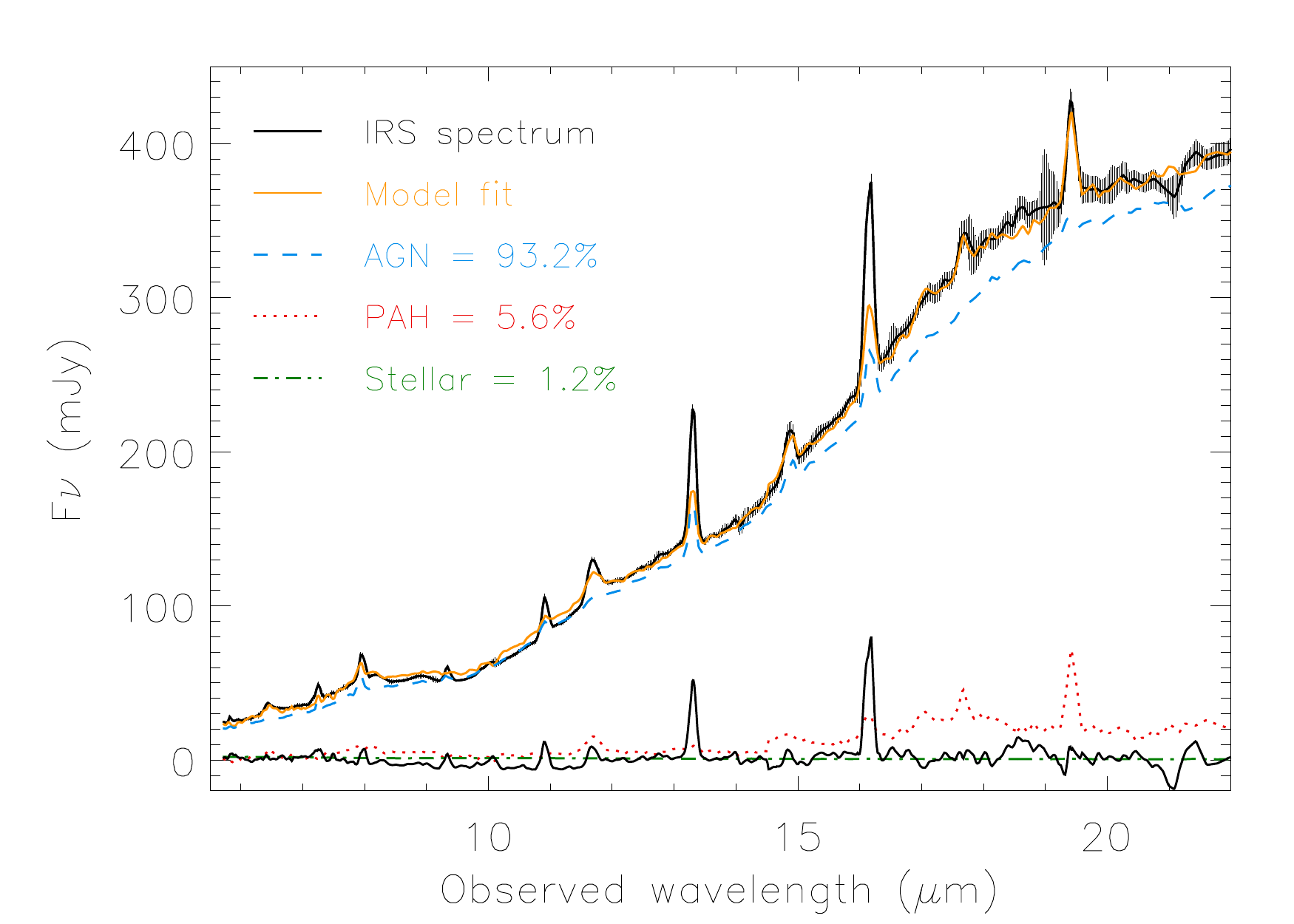}
   \caption{IRS spectrum of Mrk\,477 fitted with a combination of empirical AGN, PAH, and stellar emission templates. The different components are indicated with different colors and symbols, and the residuals from the fit are shown as a solid black line at around 0 in flux density.}
              \label{figA1}
   \end{figure}

We used the {DeblendIRS} tool \citep{2015ApJ...803..109H} to decompose the IRS spectrum of Mrk\,477, in the range 5--22 $\mu$m, into different components. The fit was done using a linear combination of empirical AGN, PAH, and stellar emission templates that minimizes the $\chi^2$. The AGN templates correspond to IRS spectra from CASSIS of different types of active galaxies, the PAH templates to star-forming or starburst galaxies, and the stellar templates to elliptical and S0 galaxies with weak or absent PAH features. From the fit shown in Fig.~\ref{figA1}, we find that the AGN, PAH, and stellar emission contributions to the mid-IR continuum are 93.2\%, 5.6\%, and 1.2\%, respectively. This combination of templates produces a $\chi^2$=2.876.

\end{appendix}

\end{document}